\documentclass[12pt]{article}
 \csname
@addtoreset\endcsname{equation}{section}

\def\a{\alpha} \def\ad{\dot{\a}} 
\def\hA{{\widehat A}}

\def\b{\beta}  \def\bd{\dot{\b}} 
\def\c{\gamma} \def\cd{\dot{\c}}
\def\C{\Gamma}
\def\d{\delta} \def\dd{\dot{\d}}

\def\e{\epsilon}

\def\F{\Phi}

\def\k{\kappa}
\def\l{\lambda}
\def\L{\Lambda}
\def\m{\mu}
\def\n{\nu}
\def\r{\rho}
\def\s{\sigma}

\def\t{\tau}
\def\th{\theta} 

\def\x{\xi}

\def\O{\Omega}
\def\o{\omega}
\def\pb{{\bar\pi}}

\def\cA{{\cal A}}

\def\cN{{\cal N}}

\def\cV{{\cal V}}

\def\cA{{\cal A}}
\def\hcA{\widehat\cA}
\def\hF{\widehat{\F}}

\def\bp{{\bar \pi}}
\def\yb{{\bar y}}
\def\zb{{\bar z}}

\def\ra{\rightarrow}

\let\la=\label
\let\bm=\bibitem
\def\nn{\nonumber}
\newcommand{\eq}[1]{(\ref{#1})}

\newcommand{\w}[1]{\\[0.#1cm]}
 
\def\be{\begin{equation}}
\def\ee{\end{equation}}
\def\bea{\begin{eqnarray}}
\def\eea{\end{eqnarray}}
\def\ba{\begin{array}}
\def\ea{\end{array}}
\def\se{\;\;=\;\;}

\def\mx#1#2#3#4{\left#1\begin{array}{#2} #3 \end{array}\right#4}

\def\ft#1#2{{\textstyle{{\scriptstyle #1}
\over {\scriptstyle #2}}}}

\begin{document}

\hfill{CTP-TAMU-09/02}
\\[-20pt]

\hfill{UU-02-02}
\\[-20pt]

\hfill{hep-th/0205132}

\hfill{\today}

\vspace{20pt}
\begin{center}


{\Large \bf  Analysis of Higher Spin Field Equations}\\

\vspace{0.2cm} {\Large\bf in Four Dimensions}


\vspace{30pt}
{\sf\large E. Sezgin}\\[5pt]

{\it\small Center for Theoretical Physics, Texas A\&M University,
College Station, TX 77843, USA}\vspace{15pt}

{\sf\large P. Sundell}\\[5pt]
{\it\small Department for Theoretical Physics, Uppsala
Universitet, Sweden}


\vspace{60pt} {\bf Abstract}\end{center}

The minimal bosonic higher spin gauge theory in four dimensions
contains massless particles of spin $s=0,2,4,..$ that arise in the
symmetric product of two spin $0$ singletons. It is based on an
infinite dimensional extension of the $AdS_4$ algebra a la
Vasiliev. We derive an expansion scheme in which the gravitational
gauge fields are treated exactly and the gravitational curvatures
and the higher spin gauge fields as weak perturbations. We also
give the details of an explicit iteration procedure for obtaining
the field equations to arbitrary order in curvatures. In
particular, we highlight the structure of all the quadratic terms
in the field equations.

\setcounter{page}{1}

\pagebreak


\section{Introduction}


The correspondence between  weakly coupled gauged supergravity in
$AdS_5$ and strongly coupled supersymmetric Yang-Mills theory
living on the four dimensional boundary of $AdS_5$ has been
intensely studied in recent years and these studies have led to
many interesting results. Much less in known, however, in the
limit in which the Yang-Mills theory is weakly coupled or even
free. Nonetheless, it is well known that in such a limit two
things happen. Firstly, the bulk supergravity is no longer a good
approximation due to the fact that the mass of the string states
become light. Secondly, the free Yang-Mills theory, by which we
mean an SU(N) Yang-Mills theory with $g_{YM}\ra 0$, admits
infinitely many conserved currents with ever increasing spin.

The above facts, together with considerations detailed elsewhere,
have led to the proposal that free SYM on the boundary of $AdS_5$
is holographic dual of a bulk theory which is an {\it interacting}
higher spin (HS) gauge theory \cite{su2,su1,us1,us2,5dv1,
5dv2,edseminar,mikhailov,us7d,holo} based on an infinite
dimensional extension of the AdS superalgebra. It has been
furthermore noted that turning on the Yang-Mills coupling constant
gives rise to a Higgs mechanism in the bulk in which the HS
symmetries are spontaneously broken down to those of the ordinary
supergravity, namely the finite dimensional AdS supersymmetry
\cite{edseminar,holo}. An interesting twist to this story has
appeared recently in \cite{gkp} where a remarkable connection
between higher spins and the dynamics of long, folded closed
strings in $AdS_5$ has been found. These developments begin to
merely scratch the surface of very interesting physics in a new
phase string/M theory which has hardly been explored so far, and
it holds the promise of exciting results.

In this paper we will neither address in detail the free CFT/HS
gauge theory correspondence, nor the construction of the
interacting, full 5D HS gauge theory; we have discussed these and
several related matters recently in \cite{holo}. Instead, here we
shall examine closely the structure of the interactions in the
simplest possible situation, namely in a minimal bosonic HS gauge
theory in 4D where the full field equations including auxiliary
fields have been known in a closed form for some time
\cite{4dv1,vr2}. However, the ingredients that go into their
formulation are not widely familiar. Moreover, it would be useful
to extract the couplings of all the physical fields involved
explicitly so that traditional field theory computations may
become feasible in order to test various aspects of the proposed
CFT/HS gauge theory correspondence. With that in mind, here we
shall analyze in greater detail than before the HS equations of
motion in the case of the minimal bosonic model in 4D. Further
work is required, however, to present them explicitly in terms of
physical fields which shows the numerical coefficients of the
couplings as well as the distribution of derivatives \cite{wip}.

The HS gauge theory model involves a star product and in many
respects is reminiscent of non-commutative gauge field theories.
The reader can find relatively detailed historical background and
references for HS gauge theories, for example, in \cite{us1}.
These theories have been primarily been developed by Vasiliev
\cite{4dv1,vr2}. In \cite{us3,us4}, we have expressed Vasiliev's
main result in a way which makes the geometrical nature of the
equations somewhat more transparent. In doing so, however, the
precise relation with Vasiliev's equations have not been made
altogether precise. This will be remedied here and we will exhibit
the precise relation in this paper.

The physical fields of the minimal bosonic model are massless
fields with spin $s=0,2,4,...$, each occurring once. For reasons
explained elsewhere (see \cite{us1}, for example, for a review),
this is precisely the spectrum which arises in the tensor product
of two spin 0 singleton representations of the AdS group $SO(3,2)$
\cite{ff}. Extensions of the model to the supersymmetric case as
well as higher dimensions show that this bosonic spectrum is
always contained as a minimal bosonic truncation
\cite{us1,us2,us7d,holo} (which is consistent in 4D where the
interactions are known). Given also the fact that the minimal
bosonic model already exhibits many features of the higher
supersymmetric models, it provides an ideal ground for studying
the basic properties of HS gauge theory.

The notion of curvature expansion has been introduced by Vasiliev
who studied the expansion for a certain 4D HS gauge theory in up
to second order in \cite{v2nd}. Here we develop further the
generally covariant expansion scheme \cite{ssc}, in which the
gravitational gauge fields are treated exactly while all
curvatures, higher spin gauge fields and the physical scalar field
are treated as weak fields. In particular, we describe a scheme
for solving the highly complicated constraint equations for the
auxiliary fields to any order in the weak fields.

An action, with or without auxiliary fields, which reproduces the
full field equations is not known. It should nevertheless be
possible to use the weak field expansion scheme to build an action
for the physical fields, including the boundary terms which are
important for computing bulk-to-boundary amplitudes. It would also
be interesting to exhibit the form of the scalar potential. We
hope to address these issues elsewhere.

This paper is organized as follows. In Section 2 we define the
minimal bosonic model and describe its full field equations
including auxiliary fields in an expansion in terms of the scalar
field, the generalized Weyl tensors and their derivatives. In
Section 3 we show how the minimal bosonic model can be obtained
from the $\cN=2$ model of Vasiliev \cite{4dv1} by consistent
truncation. In Section 4, we elaborate on the Lorentz
transformation of the master fields ensuring that the component
fields transform in an appropriate manner \cite{vr2}. In Section
5, we describe a weak field expansion scheme for eliminating the
auxiliary fields and obtaining the interactions among the physical
fields to any desired order. In Section 6, we take a closer look
at the field equations including second order terms, and exhibit
their structure in considerable detail. In Section 7 we comment on
our results, holography and some open problems.


\section{The Minimal Bosonic Model and Curvature Expansion}


The minimal bosonic theory in 4D is based on the higher spin
algebra $hs_2(1)$ \cite{kv2}, which we shall refer to in this
paper as $hs(4)$. This algebra is realized in terms of oscillators
obeying the following algebra

\be y_\a\star y_\b=y_\a y_\b+i\e_{\a\b}\ ,\quad y_\a\star
\yb_{\ad}= y_\a \yb_{\ad}\ ,\quad (y_\a)^\dagger=\yb_{\ad}\ ,
\la{y}\ee

where $y_\a$ $(\a=1,2)$ is a Weyl spinor which is a Grassmann even
generator of a Heisenberg algebra. The $\star$ denotes the
associative product between oscillators. The products on the right
hand sides are Weyl ordered, so that for example $y_\a y_\b=y_\b
y_\a$. Using the above contraction rules it is straightforward to
compute $f\star g$ where $f(y,\yb)$ and $g(y,\yb)$ are two arbitrary Weyl 
ordered
polynomials of oscillators (see \eq{star} below). The resulting $\star$-algebra

is
associative.

The algebra $hs(4)$ consists of arbitrary Grassmann even and
anti-hermitian polynomials $P(y,\yb,\th)$ that are sums of
monomials of degree $4\ell+2$ where $\ell=0,1,2,...$, which will
be referred to as the level index. The Lie bracket between $P,Q\in
hs(4)$ is given by $[P,Q]_\star$. Thus, denoting by $P^{(\ell)}$
an $\ell$th level monomial, the commutation relations have the
schematic form:

\be [P^{(\ell_1)},P^{(\ell_2)}]_\star =\sum_{|\ell_1-\ell_2|\leq
\ell \leq \ell_1+\ell_2} P^{(\ell)}\ .\la{ell}\ee

In particular, the zeroth level of $hs(4)$ is the maximal finite
subalgebra $SO(3,2)\simeq Sp(4)$ whose generators schematically
take the form

\be M_{\a\bd}= y_{\phantom{\bd}\!\!\!\a} \yb_{\bd}\ ,\quad\ M_{\a\b}=y_\a 
y_\b\ ,
\quad\ M_{\ad\bd}=\yb_{\ad}\yb_{\bd}\ . \la{ads}\ee

A generator $P^{(\ell)}$ in the $\ell$th level of $hs(4)$ can be
expanded as

\be P^{(\ell)}(y,\yb,\th)=
\sum_{\footnotesize\ba{c}m+n\\=4\ell+2\ea}{1\over m!\,n!}~
\yb^{\ad_1}\cdots \yb^{\ad_m} y^{\b_1}\cdots y^{\b_n}
P_{\ad_1\dots\ad_m\,\b_1\dots\b_n}\ .\la{p4}\ee

The spins of the components are given by $s=\ft12(m+n)$. The
reality properties follow from $P^\dagger=-P$. It is useful to
summarize the conditions that define $P$ as

\be \tau(P)=-P\ ,\qquad P^\dagger=-P\ , \ee

where the $\tau$-map is defined as

\be \tau(y)=iy\ ,\qquad \tau(\yb)=i\yb\ . \la{tau}\ee

The $\t$-map obeys $\t(f\star g)=\t(g)\star\t(f)$ which makes it an anti-
involution of
the associative $\star$-algebra.

The basic building blocks of the HS gauge theory based on this
algebra are the master 1-form $A(x;y,\yb)=dx^\m A_\m(x;y,\yb)$, where $x^\m$
are
the coordinates of the 4D spacetime, and
a master 0-form $\Phi(x;y,\yb)$ which satisfy the conditions

\be \tau(A_\m)= -A_\m\ ,\quad (A_\m)^\dagger =-A_\m\ ,\qquad
{\tau}(\Phi)={\bar\pi}(\Phi)\ ,\quad \Phi^{\dagger}=\pi(\Phi)\
,\la{hf1}\ee

where the maps $\pi$ and $\bar{\pi}$ maps are defined as

\bea \pi\,(y_\a)&=&-y_\a \ ,\qquad \bp\,(y_\a)\se y_\a\ ,\nn\w2
\pi\,(\yb_{\ad})&=& \yb_{\ad}\ ,\qquad\ \ \, \bp\,(\yb_{\ad})\se
-\yb_{\ad}\ . \la{pz1} \eea

These maps obey $\pi(f\star g)=\pi(f)\star\pi(g)$ and $\bp(f\star g)=\bp(f)
\star\bp(g)$
which make them involutions of the associative $\star$-algebra.
The rationale behind \eq{pz1} is explained in detail, for
example, in \cite{ssc}. The conditions on $A_\m$ define the adjoint
representation of the higher spin algebra $hs(4)$ while the
conditions on $\Phi$ define a quasi-adjoint representation of
$hs(4)$. The $\pi$-twist is essential for obtaining the physical
spectrum of states. The conditions on $A_\m$ ensure that the
gauge fields contained in it have even integer spins $s=2,4,6,...$
define an $hs(4)$-valued spacetime one-form and a spacetime
zero-form in a certain quasi-adjoint representation of $hs(4)$
\cite{ssc}. The component field expansion of the master fields
reads\footnote{From the condition $\tau(\Phi)=\bp(\Phi)$ it
follows that $\bp(\Phi)=\pi(\Phi)$, so that $\t(\F)=\bp(\F)$ and
$\t(\F)=\pi(\F)$ give rise to equivalent representations.}:

\bea A_\mu(x;y,\yb)&=& {1\over 2i} \sum_{m+n=2\ {\rm mod}\ 4}
{1\over m!n!} \yb^{\ad_1}\cdots\yb^{\ad_m} y^{\a_1}\cdots y^{\a_n}
A_{\m \a_1\dots\a_n\ad_1\dots\ad_m}(x)\ ,\la{aexp}\\[10pt]
\Phi(x;y,\yb)&=& \sum_{|m-n|=0\ {\rm mod}\ 4} {1\over m!n!}
\yb^{\ad_1}\cdots\yb^{\ad_m} y^{\a_1}\cdots y^{\a_n} \Phi_{
\a_1\dots\a_n\ad_1\dots\ad_m}(x)\ .\la{Phiexp}\eea

To describe the field equations in 4D spacetime one introduces an
auxiliary set of coordinates $Z=(z^\a,\bar{z}^{\ad})$ which are
Grassmann even spinors that are non-commutative in nature
\cite{4dv1}. To be more precise, in addition to \eq{y}, one
postulates the $\star$-commutation rules

\be z_{\a}\star z_{\b}= z_{\a}z_{\b}-i\,\e_{\a\b}\ ,\qquad
y_{\a}\star z_{\b}= y_{\a}z_{\b}-i\,\e_{\a\b}\ ,\qquad z_{\a}\star
y_{\b}= z_{\a}y_{\b}+i\,\e_{\a\b}\ ,\la{zy} \ee

and their complex conjugates with $(z_\a)^\dagger=\zb_{\ad}$. The
resulting associative $\star$-product between two arbitrary
Weyl-ordered polynomials 
$\widehat f(y,\yb;z,\zb)$ and $\widehat g(y,\yb;z,\zb)$ is defined as

\be \widehat f*\widehat g=\widehat f \exp\left[
i\left({\overleftarrow{\partial}
\over
\partial z_\a} + {\overleftarrow{\partial}\over \partial
y_\a}\right) \left({\overrightarrow{\partial}\over \partial z^\a}
- {{\overrightarrow\partial}\over \partial y^\a}\right)
+i\left({\overleftarrow{\partial}\over \partial {\zb}_{\ad}}
-{\overleftarrow{\partial}\over \partial {\yb}_{\ad}}\right)
\left({\overrightarrow{\partial\over \partial {\zb}^{\ad}}} +
{\overrightarrow{\partial}\over \partial {\yb}^{\ad}}\right)
\right]\widehat g\ . \la{star}\ee

Here and in the rest of the paper, hatted quantities depend on
$(z,\zb)$. One then considers extensions $\hA$ and $\hF$ of the
basic master fields to the $(x;z,\zb)$ space such that
$A=\hA|_{Z=0}$ and $\F=\hF|_{Z=0}$. One then imposes constraints
on curvatures $\widehat F$ and $\widehat D\hF$ (see \eq{c1},
\eq{c2}, \eq{hatF} and \eq{DPhi} below) whose $(z,\zb)$-components
determine the $(z,\zb)$-dependence of $\hA$ and $\hF$ in terms of
$A$ and $\F$. The remaining, non-trivial curvature constraints are
contained in the reduced curvature constraints $\widehat F|_{Z=0}$
and $\widehat D\hF|_{Z=0}$ (see \eq{cn} and \eq{bn} below), which
are integrable by construction and one can show that they contain
the physical field equations of the HS gauge theory. It is
important to note that since $(z,\zb)$ are non-commutative the
reduced constraints contain highly non-trivial interactions even
though the original constraint in $(x;z,\zb)$ space has a simple
form.

The extensions of the basic master fields are a $0$-form
$\widehat{\Phi}$ and a $1$-form

\be \widehat{A}=dx^\mu \widehat{A}_\mu + dz^\a \widehat{A}_\a+
d\zb^{\ad}\widehat{A}_{\ad}\ . \ee

The component fields of $\widehat{A}$ and $\widehat{\Phi}$ are
functions of $(x;z,\zb)$ with $(y,\yb)$ expansions obeying the
conditions

\be {\tau}(\hA)= -\hA\ ,\quad \hA^\dagger =-\hA\ ,\quad\quad
{\tau}(\widehat\Phi)={\bar\pi}(\hF)\ ,\quad
\hF^{\dagger}=\pi(\hF)\ ,\la{hf2}\ee

where the anti-involution $\t$ and involution $\pi$ and $\bar{\pi}$ are defined

by
\eq{tau}, \eq{pz1}, and

\bea \pi\,(z_{\a})&=&-z_{\a}\ ,\qquad \bp\,(z_\a)\se z_\a\
,\qquad\ \ \, \t\,(z_\a)\se -i\,z_{\a}\ ,\nn\w2
\pi\,(\zb_{\ad})&=& \zb_{\ad}\ ,\qquad\ \ \, \bp\,(\zb_{\ad})\se
-\zb_{\ad}\ ,\qquad \t\,(\zb_{\ad})\se-i\,\zb_{\ad}\ , \la{pz2}
\eea

and that $d(\t(\widehat f))=\t (d(\widehat f))$, idem $\pi$ and $\bp$.
Thus, the hatted fields are given as expansions order by order in
$z$ and $\zb$, with expansion coefficients that in turn are
functions of $x$ with $(y,\yb)$ expansions obeying \eq{hf2}.
These conditions are engineered such that at $Z\equiv(z,\zb)=0$,
the pulled-back components

\be A_{\mu}=\widehat{A}_\mu|_{Z=0}\ ,\qquad
\Phi=\widehat{\Phi}|_{Z=0}\la{ic}\ee

can be identified with the basic master defined in \eq{hf1}.

The general method for formulating master curvature constraints
giving rise to full higher spin field equations in $D=4$ has been
developed by Vasiliev \cite{4dv1}. In particular, a non-minimal
bosonic model whose spectrum contains massless physical fields
with spin $s=0,1,2,...$, each occurring once, has been described
in \cite{vr1}. Here we eliminate the odd spins by imposing
conditions given in \eq{hf1}, thereby obtaining the minimal
bosonic model. These conditions are formally the same as those
imposed on the basic master fields of the minimal $\cN=8$ model,
which was studied in great detail in \cite{us3,us4}. Naturally,
the master curvature constraints of the minimal bosonic model also
take the same form as those of the $\cN=8$ model. Of course, the
fermionic oscillators of the $\cN=8$ model are to be dropped and
the $SO(8)$ chirality operator $\C$ occurring in that model is to
be replaced by unity. Thus the master curvature constraints giving
rise to the full field equations for the minimal bosonic theory
are

\bea \widehat{F}&=& \ft{i}4 dz^\a\wedge dz_\a
\cV(\widehat{\Phi}\star \k) + \ft{i}4 d\zb^{\ad}\wedge d\zb_{\ad}
\bar{\cV}(\widehat{\Phi}\star \bar{\k})\ ,\la{c1}\w4
\widehat{D}\widehat{\Phi}&=& 0\ ,\la{c2}\eea

where the curvatures are defined as

\bea \widehat{F}&=& d\widehat{A}+\widehat{A}\star\widehat{A}\ ,
\la{hatF}\w2
\widehat{D}\widehat{\Phi}&=&d\widehat{\Phi}+\widehat{A}\star
\widehat{\Phi}-\widehat{\Phi}\star \bar{\pi}(\widehat{A})\ .\la{DPhi}\eea

and the operators $\kappa,{\bar\kappa}$ as

\be \k=\exp(iy^\a z_\a)\ ,\quad\quad
\bar{\k}=\k^\dagger=\exp(-i\yb^{\ad}\zb_{\ad})\ .\ee

The quantity $\cV(X)$ in \eq{c1} is a $\star$-function, $\cV(X)=
b_0+b_1X+b_2 X\star X+\cdots$, with its complex conjugate $\cV(X)$
defined by $\bar{\cV}(X)=(\cV(X^\dagger))^\dagger$. The
requirement that the theory admits proper local Lorentz
transformations \cite{vr2}, as we shall discuss in detail in
Section 4, reduces the amount of freedom in $\cV$. The possibility
of redefining $\hF$ while preserving the basic structure of the
curvature constraint \eq{c1} \cite{4dv1,us4} leads to further
reductions in the number of free parameters in $\cV$. The result
is that $\cV$ is an odd function,

\be \cV(X)=b_1X+b_3X\star X\star X+\cdots\ ,\la{odd}\ee

which is defined modulo the field redefinitions $\hF\ra F(\hF)$
where $F$ is a real and odd $\star$-function, i.e.

\be \cV(X)\sim \cV(F(X))\ ,\qquad F(X)=-F(-X)\ ,\qquad
(F(X))^\dagger=F(X^\dagger)\ .\la{cvredef}\ee

As a result there is no loss of generality in setting

\be |b_1|=1\ ,\qquad b_1=e^{i\th_1}\ .\la{be1}\ee

We shall see that even the simplest choice $\cV(X)=b_1 X$ leads to
highly complicated interactions which are higher order in
derivatives. Adding higher order terms to $\cV(X)$ lead to
modifications of the higher order interactions, which are, of
course, consistent with HS gauge invariance. The free parameters
in $\cV$ are therefore genuine interaction ambiguities. It has
been proposed that these parameters can be fixed upon comparison
with a underlying singleton theory \cite{holo}.

It is important to note that the constraints \eq{c1} and \eq{c2}
are integrable and have the symmetry

\bea \d
\widehat{A}&=&d\widehat{\e}+[\widehat{A},\widehat{\e}]_{\star}\
,\la{cgs1}\\[10pt]
\d\widehat{\Phi}&=&\widehat{\Phi}\star\bar\pi(\widehat{\e})-\widehat{\e}
\star\widehat{\Phi} \ ,\la{cgs}\eea

where $\widehat{\e}$ obeys the same conditions as those imposed on
$\hA$ in \eq{hf2}, that is

\be \tau(\widehat{\e})=-\widehat{\e}\ ,\qquad
(\widehat{\e})^\dagger=-\widehat{\e}\ .\la{he1}\ee

The components of the constraints \eq{c1} and \eq{c2} which have
at least one $z^\a$ or $\zb^{\ad}$ index determine the
$Z$-dependence of $\widehat{A}$ and $\widehat{\Phi}$ in terms of
the initial conditions \eq{ic} up to a gauge transformation. By
assuming that $\hF$ and $\hA$ can be expanded in powers of $\Phi$,
which contains curvatures and the scalar field, we can obtain the
$Z$-dependence of the master fields by making the following
ansatz:

\be \widehat{\Phi}=\sum_{n=1}^\infty \widehat{\Phi}^{(n)}\ ,\quad
\widehat{A}_{\a}=\sum_{n=0}^\infty \widehat{A}_\a^{(n)}\ ,\quad
\widehat{A}_{\mu}= \sum_{n=0}^\infty \widehat{A}_\mu^{(n)}\
,\la{ces}\ee

where $\widehat{\Phi}^{(n)}$ ($n=1,2,3,...$), $\widehat{A}^{(n)}_\a$
($n=0,1,2,...$) and $\widehat{A}_\m^{(n)}$ ($n=0,1,2,...$) are
funcionals which are $n$th order in $\Phi$ and which obey the initial
conditions

\bea \widehat{\Phi}^{(n)}|_{Z=0}&=&\mx{\{}{ll}{\Phi\ ,&n=1\\0\
,&n=2,3,...}{.}\la{phiic} \w2 \widehat{A}_\m^{(n)}|_{Z=0} &=&
\mx{\{}{ll}{A_\m\ ,&n=0\\0\ ,&n=1,2,3,...}{.}\la{amic}\
.\la{aaic}\eea

>From $\widehat F^{(0)}_{\a\b}=0=\widehat
F^{(0)}_{\a\bd}$ it follows that $\hA^{(0)}_\a$ is a pure gauge
artifact which we can eliminate by imposing the gauge condition

\be \hA^{(0)}_\a=0\ .\la{gauge}\ee

Next, the constraints $\widehat{F}_{\a\bd}=0$,
$\widehat{F}_{\a\b}=-\ft{i}2\e_{\a\b}\cV(\hF\star\k)$ and
$\widehat D_\a\hF=0$ can be solved in the $n$th order ($n\geq 1$) as

\bea \hF^{(1)}&=&\F(y,\yb)\ ,\\
\hcA^{(1)}_\a&=& \partial_\a\widehat \x^{(1)}-{i b_1\over 2} z_\a
\int_0^1 tdt~
 \F(-tz,\yb)\kappa(tz,y)\ ,\la{ah}\eea

and ($n\geq 2$):

\bea \hF^{(n)}&=& z^\a\sum_{j=1}^{n-1}\int_0^1 dt \Bigg(
\hF^{(j)}\star
\pb(\widehat{A}^{(n-j)}_\a)-\widehat{A}^{(n-j)}_\a\star
\hF^{(j)}\Bigg)_{z\ra tz,\zb\ra t\zb}\nn \w2 &&+~
\zb^{\ad}\sum_{j=1}^{n-1}\int_0^1 dt \Bigg( \hF^{(j)}\star
\pi(\widehat{A}^{(n-j)}_{\ad})-\widehat{A}^{(n-j)}_{\ad}\star
\hF^{(j)}\Bigg)_{z\ra tz,\zb\ra t\zb}\ ,\la{it1}\w2
\widehat{A}^{(n)}_\a&=&\partial_\a\widehat\x^{(n)}+z_\a \int_0^1
tdt\Bigg(-\ft{i}2 \cV^{(n)}(\hF\star\kappa) +
\sum_{j=1}^{n-1}\widehat{A}^{(j)\b}\star
\widehat{A}^{(n-j)}_\b\Bigg)_{z\ra tz,\zb\ra t\zb} \nn\w2
&&+\zb^{\bd}\sum_{j=1}^{n-1}\int_0^1 tdt \left[
\widehat{A}^{(j)}_{\a},\widehat{A}^{(n-j)}_{\bd}\right]_{* (z\ra
tz,\zb\ra t\zb)}\ ,\la{it2} \eea

where $\cV^{(n)}(\widehat{\Phi}\star\k)$ is the $n$th order term
of $\cV(\widehat{\Phi}\star\k)$ in the curvature expansion scheme
\eq{ces}. We emphasize that in \eq{ln}, \eq{it1} and \eq{it2} the
replacements $(z,\zb)$ by $(tz,t\zb)$ are to be made {\it after}
the $\star$-products are carried out. The integration functions
$\widehat \x^{(n)}$ are gauge artifacts which can be eliminated by means of
$\F$-dependent gauge transformations. We therefore impose the
gauge conditions \cite{us3}

\be \widehat\x^{(n)}=0\ ,\qquad n=1,2,\dots\ \la{gauge1}\ee

The gauge conditions \eq{gauge} and \eq{gauge1} are left invariant
by $Z$-independent, and therefore $hs(4)$-valued, gauge
transformations (which in general may be $\F$-dependent).

>From the constraint $\widehat{F}_{\m\a}=0$ in \eq{c1} it follows
that

\be \widehat{A}_\mu ={1\over
1+\widehat{L}^{(1)}+\widehat{L}^{(2)}+\widehat{L}^{(3)}+\cdots}
~A_\mu\ ,\ee\smallskip

where the linear operators $\widehat{L}^{(n)}$ are defined by

\bea \widehat{L}^{(n)}(\widehat f)= -i\int_0^1{dt\over t}
&\Bigg[&\widehat{A}^{\a(n)}\star \left({\partial \widehat f\over \partial
z^\a} -{\partial \widehat f\over \partial y^\a}\right) +\left({\partial
\widehat f\over
\partial z^\a} +{\partial \widehat f\over \partial y^\a}\right) \star
\widehat{A}^{\a(n)} \la{ln}\w2 &+&\widehat{A}^{\ad(n)}\star
\left({\partial \widehat f\over \partial {\zb}^{\ad}} +{\partial \widehat
f\over
\partial {\yb}^{\ad}} \right) +\left({\partial \widehat f\over \partial
{\zb}^{\ad}} -{\partial \widehat f\over \partial {\yb}^{\ad}}\right) \star
\widehat{A}^{\ad(n)}\Bigg]_{z\ra tz,\zb\ra t\zb} \la{lin}\eea

Note that diffeomorphism invariance in $x^\m$ requires
$\hA_\m$ to be linear in $A_\m$ or $\partial_\m\F$. However,
the gauge conditions \eq{gauge} and \eq{gauge1}
imply that the
$\partial_\m\F$ terms cancel identically \cite{ssc}.

Finally, having solved the $Z$-space part of \eq{c1} and \eq{c2},
the remaining constraints $\widehat{F}_{\m\n}=0$ and
$\widehat{D}_\m \widehat{\Phi}=0$ yield the following spacetime
field equations\footnote{The integrability of \eq{c1} and \eq{c2} implies
that if $\widehat{F}_{\m\n}|_{Z=0}=0$ and
$\widehat{D}_\m \widehat{\Phi}|_{Z=0}=0$ then $\widehat{F}_{\m\n}=0$ and
$\widehat{D}_\m \widehat{\Phi}=0$.}:

\bea F_{\m\n} &=&-2\sum_{n=1}^\infty\sum_{j=0}^{n} \Bigg(
\widehat{A}_{[\m}^{(j)}\star
\widehat{A}_{\n]}^{(n-j)}\Bigg)_{|_{Z=0}}\ , \la{cn}\w2 D_\m \Phi
&=& \sum_{n=2}^\infty  \sum_{j=1}^{n}
\Bigg(\widehat{\Phi}^{(j)}\star \bar
\pi(\widehat{A}^{(n-j)}_\m)-\widehat{A}^{(n-j)}_\m\star
\widehat{\Phi}^{(j)} \Bigg)_{|_{Z=0}}\ , \la{bn} \eea

where

\be F=dA+A\star A\ ,\quad\quad  D\Phi=d\Phi+A\star
\Phi-\Phi\star\bar{\pi}(A)\ .\ee

It is important that \eq{cn} and \eq{bn} are integrable equations.
As such they are invariant under gauge transformations whose form
follow uniquely by functional variation of \eq{cn} and \eq{bn}, as
is discussed in more detail at the end of Section 5. Equivalently,
these symmetries can be described as the residual $hs(4)$-valued
gauge transformations discussed above.

Before we proceed further to analyze in more detail the structure
of these equations, it is useful to compare the constraints
\eq{c1} and \eq{c2} with those of Vasiliev \cite{4dv1}.


\section{The Minimal Bosonic Theory from Vasiliev's $\cN=2$ Model}


The original formulation of the HS field equations in $D=4$
\cite{4dv1} is based on the $\cN=2$ higher spin algebra
$hu(1,1|4)$ \cite{kv2}, which is obtained by adding two extra
elements $k$ and $\bar k$, which are known as Kleinian operators,
with the following properties\footnote{More recently, $\cN\geq 1$
models formulated without Kleinian operators have been discussed
in \cite{n124}.}

\be k^2=\bar k^2=1\ ,\quad [k,\bar k]=[k,\bar
y_{\ad}]=\{k,y_\a\}=[\bar k,y_\a]=\{\bar k,\bar y_{\ad}\}=0\
.\la{kr}\ee

The corresponding enlarged master fields, which we shall denote here
by $B_\m$ and $\Psi$, were taken in \cite{4dv1} to obey the
reality conditions

\be (B_\m)^\dagger=-B_\m\ ,\quad (\Psi)^\dagger=\Psi\ .\ee

The maximal finite subalgebra of $hu(1,1|4)$ is $OSp(2|4)\times U(1)$ with 
supercharges and
$U(1)_R$ generator given by \cite{vr2}:

\be \quad Q^1_\a=y_\a\ ,\quad Q^2_{\a}=k \bar k y_{\a}\
,\quad J=k\bar k\ ,\la{n2}\ee

and the extra $U(1)$ generator given by $1$.

The model is then extended into the $Z$-space described in the
previous section, with analogs of \eq{kr} holding also
between the Kleinians and $(z_\a,\bar z_{\ad})$. The basic master
constraints first given in \cite{4dv1} are equivalent to the curvature
constraint

\bea d\widehat B+\widehat B \star \widehat B &=& \ft{i}4 dz^2
\cV(\widehat \Psi)\star (k\kappa) + \ft{i}4 d\bar z^2 \bar
\cV(\widehat \Psi)\star (\bar k\bar \kappa)\ ,\la{kmc1}\w2
d\widehat \Psi+[\widehat B,\widehat \Psi]_{\star}&=&0\
,\la{kmc2}\eea

where $\widehat B=dx^\m \widehat B_\m+dz^\a \widehat B_\a +d\bar
z^{\ad}\widehat B_{\ad}$ and $\widehat \Psi$ are extensions of
$B$ and $\Psi$ obeying

\be (\widehat B)^\dagger = -\widehat B\ ,\quad
(\widehat \Psi)^\dagger=\widehat \Psi\ .\ee

The dependence of the master fields on the Kleinians results in
extra auxiliary as well as physical fields, compared with the
minimal bosonic model. Letting

\bea \widehat B_\m(k,\bar k)&=&\widehat B_{0\,\m}(k,\bar
k)+\widehat B_{1\,\m}(k,\bar k)\ ,\quad \widehat B_{r~
\m}(-k,-\bar k)=(-1)^r \widehat B_{r\,\m }(k,\bar k)\ ,\w2
\widehat \Psi (k,\bar k)&=&\widehat \Psi_{0}(k,\bar k)+\widehat
\Psi_{1}(k,\bar k)\ ,\quad \widehat \Psi_{r}(-k,-\bar k)=(-1)^r
\widehat \Psi_{r}(k,\bar k)\ ,\eea

one can show that the initial conditions $B_{1\m}$ and $\Psi_{0}$
contain no physical fields \cite{4dv1}, i.e. all components of
these fields are auxiliary at the linearized level. Moreover,
from the structure of \eq{kmc1} and \eq{kmc2} it follows that

\be \widehat B_{1\,\m}=0\ , \quad\quad \widehat \Psi_{0}=0\ ,
\la{t1}\ee

is a consistent truncation provided that $\cV(X)$ obeys \eq{odd}.
The remaining master fields $B_{0\,\m}$ and $\Psi_{1}$ contain
physical fields which fill up $OSp(2|4)$ supermultiplets.

This model can be truncated further to the minimal bosonic model by first 
imposing

\be \t( B_{0\,\m})=-B_{0\,\m}\ ,\quad \t( \Psi)= \Psi\
,\la{tbpsi}\ee

where $\t(k)=k$ and $\t(\bar k)=\bar k$, and then identifying

\be k\bar k\simeq 1\ .\la{kksim1}\ee

>From \eq{tbpsi} it follows that

\be B_{0\,\m}=A'_\m+k\bar k A''_\m\ ,\quad\quad \Psi_{1}=\chi k +
\bar \pi(\chi^\dagger)\bar k \ , \ee

where $A'_\m$ and $A''_\m$ obeys \eq{hf1} and $\chi$ is
unconstrained. Note that the generators \eq{n2} are dropped
from the expansion of  $A'_\m$ and $A''_\m$. Using \eq{kksim1}
we can identify the master fields $A_\m$ and $\Phi$ obeying \eq{hf1}
as

\be B_{0\m}\sim A'_\m+A''_\m\equiv A_\m\ ,\quad\quad \Psi_1
~k\sim\Psi_1~\bar k\sim \chi + \bar \pi(\chi^\dagger)\equiv \F\ .\ee

We conclude that the $\cN=2$ master constraints \eq{kmc1} and
\eq{kmc2} of the $hu(1,1|4)$ model can be truncated consistently
to those of the minimal bosonic $hs(4)$ model by means of \eq{t1},
\eq{tbpsi} and \eq{kksim1} provided that $\cV$ obeys \eq{odd}.


\section{ Lorentz Transformation of the Master Fields}


Turning to the minimal 4D bosonic HS gauge theory described in
Section 2, an analysis of the constraints \eq{cn} and \eq{bn} to
derive the full field equations requires, among other things, an
identification of component fields that transform properly under
Lorentz transformation. The fact that the component fields of the
naive $y$ and $\yb$ expansion of $A_\m$ do not transform as
conventional Lorentz tensors was first pointed out in \cite{vr2}
where the field redefinition which defines the proper Lorentz
tensors in $A_\m$ was first given.

Following \cite{vr2}, we begin by extending the Lorentz generator $M_{\a\b}$
defined in \eq{ads} to

\be \widehat{M}_{\a\b}= y_\a y_\b - z_\a z_\b \ .\ee

Commutation with $\widehat{M}_{\a\b}$ generates Lorentz rotations
on $y_\a$ and $z_\a$. The associated Lorentz parameter is

\be {\widehat \e}_0(\L) = {1\over 4i}
\L^{\a\b}\widehat{M}_{\a\b}-{\rm h.c.} \ .\ee

Next, it is important to examine how $\hA_\a$ transforms under
this Lorentz transformation, especially in view of the fact that
this master field carries a free spinorial index and that the
gauge choice \eq{gauge} is to be maintained. Under ${\widehat
\e}_0(\L)$ transformations

\be \d\hA_\a = [{\widehat\e}_0(\L), \hA_\a]_* +{1\over
2i}\L_\a{}^\b z_\b\ . \la{Lorentz1}\ee

This shows that the free spinorial index of $\hA_\a$ does not
rotate properly under ${\widehat \e}_0(\L)$ transformations, and
that as a result the transformed field no longer satisfy
\eq{gauge}. Therefore these transformations need to be modified.
To do so, it is convenient to define \cite{4dv1}

\be \widehat{S}_\a = z_\a-2i\widehat{A}_\a\ ,\ee

and rewrite \eq{Lorentz1} as

\be \d \hA_\a = [{\widehat\e}_0(\L), \hA_\a]_* +\L_\a{}^\b \hA_\b
+{1\over 2i} \L_\a{}^\b \widehat S_\b \ . \la{Lorentz2}\ee

The first two terms on the right hand side have the desired form
but the last term needs to be eliminated. This is achieved with
the following extra gauge transformation

\be {\widehat\e}_{\rm extra}(\L)= {1\over 8i} \L^{\a\b}\,\{
\widehat S_\a,\widehat S_\b\}_*-{\rm h.c.}\ , \la{extra} \ee

provided that $\cV(X)$ in \eq{c1} obeys \eq{odd}. This
requirement, together with the $\a$-component of the constraint
\eq{c2} which implies $S_\a\star \hF\star \k=-\hF\star \k\star
\widehat S_\a$, ensures that

\be \widehat S_\a\star \cV(\hF\star \k)+ \cV(\hF\star \k)\star
\widehat S_\a=0\ ,\la{dacv}\ee

which in turn implies that the quantity $\ft12\{\widehat{S}_\a,
\widehat{S}_\b\}_{\star}$ appearing in \eq{llhe} generates a
Lorentz rotation of $\widehat{S}_\a$ with opposite sign:

\be [\ft12\{\widehat{S}_\a,
\widehat{S}_\b\}_{\star},\widehat{S}_\c]_{\star}=-4i\widehat{S}_{(\a}\e_{\b)\c}\
.\la{sss}\ee

It follows that \eq{Lorentz2}
removes the unwanted last term in \eq{Lorentz2}
\footnote{ Note that \eq{odd} is not
needed for the integrability of the constraints \eq{c1} and
\eq{c2}, which only requires $\widehat D_{\ad}\cV(\hF\star\k)=0$ which in turn 
implies
$\widehat D_{\ad}\hF=0$ for arbitrary $\cV(X)$. As for Vasiliev's original 
$\cN=2$
model described in Section 3,
the Lorentz invariance of \eq{kmc1} and \eq{kmc2} requires
$\widehat D_\a \cV(\widehat\Psi)=0$ which holds by virtue of
$\widehat D_\a\widehat\Psi=0$ for arbitrary $\cV(X)$.}:

\be \d_{\rm extra}\hA_\a=-{1\over 2i}[\widehat
S_\a,{\widehat\e}_{\rm extra}(\L)]_{\star}=-{1\over
2i}\L_\a{}^\b\widehat S_\b\ .\la{dextra}\ee

Thus, we conclude that the appropriate local Lorentz
transformation is

\bea {\widehat\e}_L &=& {\widehat\e_0}(\L)+{\widehat\e}_{\rm
extra}(\L)\nn\w2 &=& {1\over 4i}\L^{\a\b}\left(
\widehat{M}_{\a\b}+\ft12 \{\widehat{S}_\a,
\widehat{S}_\b\}_{\star} \right)-{\rm h.c.} \nn\w2 &=& {1\over
4i}\L^{\a\b}\left( y_\a y_\b-4\widehat{A}_\a\star
\widehat{A}_\b\right) -{\rm h.c.}\ .\la{llhe} \eea

Considering the  master fields ${\widehat\Phi}$ and $\hA_\m$ as
well, all in all, we get the $\widehat\e_L(\L)$ transformations

\bea \d_{L} \hF&=&[\widehat\e_0(\L),\widehat{\Phi}]_{\star}\
,\la{ll1}\w2 \d_{L} \hA_\a &=& [\widehat\e_0(\L), {\widehat
A}_\a]_* +\L_\a{}^\b \hA_\b \ ,\la{ll2}\w2 \d_{L} \widehat{A}_\m
&=&[\widehat\e_0(\L),\widehat{A}_\m]_{\star}+{1\over
4i}\partial_\m\L^{\a\b}(\widehat M_{\a\b}+\ft12\{\widehat
S_\a,\widehat S_\b\})-{\rm h.c.}\ ,\la{ll3}\eea

In obtaining these results, we have used the fact that
${\widehat\e}_{\rm extra}(\L)\star
\widehat\Phi-\widehat\Phi\star\bar\pi({\widehat\e}_{\rm
extra}(\L))=0$ and $\widehat D_\m {\widehat\e}_{\rm
extra}(\L)=\partial_\m \L^{\a\b}\ft1{8i}\{\widehat S_\a,\widehat
S_\b\}_{\star}$ which follow from $[\widehat
S_\a,\widehat f]_{\star}=-2i\widehat D_\a \widehat f $  and the constraints 
${\widehat D}_\a
{\widehat\Phi}=0$ and ${\widehat F}_{\a \m}=0$. From \eq{ll1} and
\eq{ll3} it follows that the curvature constraints \eq{cn} and
\eq{bn} are invariant under the local Lorentz
transformations

\bea \d_L \F&=&[\e_0(\L),\Phi]_{\star}\ ,\la{ll4}\\
\d_L A_\m &=&[\e_0(\L),A_\m]_{\star}+{1\over
4i}\partial_\m\L^{\a\b}\Big[y_\a y_\b - 4
\left(\widehat{A}_{\a}\star
\widehat{A}_{\b}\right)_{Z=0}\Big]-{\rm h.c.}\ ,\la{ll5}\eea

where

\be \e_0(\L)={1\over 4i}\L^{\a\b}y_\a y_\b-{\rm h.c.}\
.\la{ll6}\ee

The transformation \eq{ll4} has the desired form. However, the
last term in the transformation \eq{ll5} shows that the components
of $A_\m$ have complicated field dependent Lorentz
transformations. We thus have to relate the components of $A_\m$
to the Lorentz connection $\o_\m^{\a\b}$, which by definition
transforms as

\be \d_L \o_\m^{\a\b}=\partial_\m \o_\m^{\a\b}+ \L^\a{}_\c
\o_\m^{\c\b}+\L^\b{}_\c \o_\m^{\c\a}\ ,\la{dlo}\ee

and a remaining set of Lorentz tensors describing the vierbein
$e_\m^{\a\ad}$ and the remaining higher spin gauge fields
(contained in $W_\m$ below). To do so one observes that from
\eq{ll2} and \eq{dlo} it follows that the quantity $\o_\m+K_\m$,
where

\bea \o_\m&=& {1\over 4i}\o_\m^{\a\b}y_\a y_\b -{\rm h.c.}\ ,\la{lor}\\
K_\m &=& i\o_\m^{\a\b}
\left(\widehat{A}_\a\star\widehat{A}_\b\right)_{Z=0}-\mbox{h.c.}
\la{k}\eea

has the following Lorentz transformation

\be \d_L(\o_\m+K_\m)=[\e_0(\L),\o_\m+K_\m]_\star+{1\over
4i}\partial_\m\L^{\a\b}\Big[y_\a y_\b - 4
\left(\widehat{A}_{\a}\star
\widehat{A}_{\b}\right)_{Z=0}\Big]-{\rm h.c.}\ .\ee

Thus $\d_L(A_\m-\o_\m-K_\m)=[\e_0(\L),A_\m-\o_\m-K_\m]_{\star}$
which means that $A_\m-\o_\m-K_\m$ can be expanded in terms of
Lorentz tensors. Hence we define

\be A_\mu= E_{\m}+ W_\mu +K_\mu\ , \ee

where $W_\m$ contains higher spin gauge fields
$W_{\m\,\a_1\dots\a_m\ad_1\dots\ad_n}$ with $m+n=6,10,...$ and
$E_\m$ contains the gravitational gauge fields:
\be E_\m = e_\m+\o_\m\ ,\quad
e_\m=\ft1{2i}e_\m^{\a\ad}y_\a\yb_{\ad}\ .\la{e} \ee

It follows that under local Lorentz transformations the vierbein
and higher spin gauge fields transform as Lorentz tensors:

\bea \d_L e_\m^{\a\ad}&=&\L^\a{}_\c e_\m^{\c\ad}+\L^{\ad}{}_{\cd}
e_\m^{\a\cd}\ ,\la{dlo2}\\ \d_L
W_{\m\,\a_1\dots\a_m\ad_1\dots\ad_n}&=& -m\L_{(\a_1}{}^\b
W_{\m\,\a_2\dots\a_m)\b\ad_1\dots\ad_n}-n\L_{\phantom{\bd}\!\!\!(\ad_1}{}^{\bd}
W_{\m\,\a_1
\dots\a_m\ad_2\dots\ad_n)\bd}\ .\la{dlo3}\eea

Now, we are ready to discuss a weak field expansion scheme which
will enable us to probe further the structure of the field
equations.


\section{Weak Field Expansion}


Let us begin by identifying the {\it physical fields}  as

\bea \Phi|_{Y=0}&=&\phi\ ,\la{ff}\w2
\left(\s^a\right)^{\a\ad}\,{\partial\over\partial
y^\a}{\partial\over\partial {\yb}^{\ad}}\,E_\m|_{Y=0}&=&e_\m{}^a\
,\nn\w2 \left(\s^{a_1}\right)^{\a_1\ad_1}\cdots
\left(\s^{a_{s-1}}\right)^{\a_{s-1}\ad_{s-1}}
\,{\partial\over\partial y^{\a_1}}\cdots {\partial\over\partial
{\yb}^{{\ad}_{s-1}}}\,W_\m|_{Y=0}&=&W_\m{}^{a_1...a_{s-1}}\ ,
\qquad s=4,6,..., \nn\eea

where $Y=(y,\yb)$. All the other components of $\Phi$, $E_\mu$ and
$W_\mu$, which we collectively refer to as the {\it auxiliary fields},
can be eliminated through the use of the constraints \eq{cn} and
\eq{bn} provided that both $\F$ and the spin $s\geq 4$ gauge fields
in $W_\m$ are treated as weak fields. Note that $\F$ contains the scalar $\phi$

and the
non-vanishing spin $s\geq 2$ curvatures (the Weyl
tensors) as well as all derivatives of these fields.

In this expansion scheme, the curvature
constraints \eq{cn} and \eq{bn} up to and including {\it first\
order} in weak fields take the form

\bea {\cal R}+ F^{(1)}(W)&=&-\left(E\star
\widehat{E}^{(1)}+\widehat{E}^{(1)}\star E \right)_{Z=0}\
,\la{lc1}\w2 d\Phi+E\star \Phi-\Phi\star \bar{\pi}(E)&=&0
,\la{lc2}\eea

where
\bea {\cal R} &=& dE+E\star E \w2
 F^{(1)}(W)&=&D_E W=dW+E\star W + W\star E\w2
 \widehat{E}^{(1)} &= & \widehat{L}^{(1)}(E)\ ,\la{E1}\eea

with  $\widehat{L}^{(1)}$ and $E$ defined in \eq{ln}and \eq{e},
respectively.  Eqs. \eq{c1} and \eq{lc2} are invariant in the
first order under the following gauge transformations\footnote{The
variation of \eq{lc1} and \eq{lc2} is proportional to the Weyl
tensor of the gravitational background (which vanishes identically
in the AdS background).}

\be \d A=d\e+[E,\e]_{\star}\ ,\quad \d\F=0\ .\la{lingt}\ee

In the first order, the parameter $b_1$ defined in \eq{odd} and \eq{be1}, which

arises on the right hand
side of \eq{lc1} through \eq{E1}, \eq{ln} and \eq{ah}, can be absorbed into a
redefinition of $\F$ \cite{us4}. We set
$b_1=1$ and from \eq{lc1} we obtain

\bea \ba{c}\mbox{Gravity}\\(s=2)\ea: && \left\{ \ba{l} {\cal
R}_{\a\b,\c\d}=\Phi_{\a\b\c\d}\ , \la{ee1}\w2 {\cal
R}_{\a\b,\cd\dd}=0\ , \w2 {\cal R}_{\a\b,\c\dd}=0\ , \ea \right.
\w4 \ba{c} \mbox{Higher spins}\\(s=4,6,...)\ea:&& \left\{ \ba{ll}
F^{(1)}_{\a\b,\c_1\dots \c_{2s-2}}=\Phi_{\a\b\c_1\dots \c_{2s-2}}\
, \la{hse1}\w3 F^{(1)}_{\a\b,\c_1\dots\c_k\cd_{k+1}\dots
\cd_{2s-2}}=0\ ,\qquad k=0,1,..,2s-3\ , \ea\right. \w4 \ba{l}
\mbox{Scalar}\\(s=0)\ea:&& \left\{ \ba{ll}
\nabla_{\a}{}^{\ad}\Phi_{\b_1\dots\b_m}{}^{
\bd_1\dots\bd_n}=i\Phi_{\a\b_1\dots\b_m}{}^{\ad\bd_1\dots\bd_n}-
imn\e_{\a(\b_1}\e^{\ad(\bd_1}\Phi_{\b_2\dots
\b_m)}{}^{\bd_2\dots\bd_n)}\ ,\la{se1}\w3 |m-n|=0\,{\rm mod}\,4\ ,
\ea\right.\phantom{aaa} \la{linscalareq}\eea

where $\nabla_{\a\ad}=(\s^\m)_{\a\ad}\nabla_\m$ is the Lorentz
covariant derivative and $F_{\a\b}=\ft12
(\s^{\m\n})_{\a\b}F_{\m\n}$, and $(\s^\m)_{\a\ad}=
(\s^a)_{\a\ad}e_a{}^\m$. The Riemann tensor
$R_a{}^b=d\o_a{}^b+\o_a{}^c\wedge \o_c{}^b$ and torsion $T^a=de^a+
e^b\wedge\o_b{}^a$ arises as

\bea {\cal R}_{\a\b,\c\d}&=&R_{\a\b,\c\d}+2\l\e_{\a(\c}\e_{\d)\b}\ ,\\
{\cal R}_{\a\b,\cd\dd}&=&R_{\a\b,\c\d}\ ,\\
{\cal R}_{\a\b,\c\dd} &=&T_{\a\b,\c\dd}\ .\eea

Equations \eq{hse1} and \eq{se1} are valid provided that the HS
gauge fields $W_\mu$ and the curvatures in $\Phi$ are small in
units of the AdS radius $R\sim |\l|^{-1/2}$. Thus, to the leading
order in this expansion scheme we can consistently set
$W_\mu=\phi=0$ in \eq{hse1} and \eq{se1}. Then \eq{ee1} describes
the full Einstein equation with a cosmological constant which is
proportional to $-|\l|$. The spin $s=2$ sector of \eq{se1}, i.e.
the components with $m=n\pm 4$, are consistent up to terms which
are second order in the spin $s=2$ Weyl tensor $\Phi_{\a\b\c\d}$.
Thus, within the expansion scheme \eq{ces}, there is a second
expansion scheme in which one expands in powers of $W_\mu$ and
$\phi$ at fixed, small $\Phi_{\a\b\c\d}$. The basic reason one can
expand around a fixed gravitational background is that the $E\star
E$ term does not act as a source for higher spins. The second
order in the double expansion scheme is obtained by including
terms in \eq{cn} and \eq{bn} which are quadratic in $\Phi$ and
$W_\m$. As a result terms quadratic in $\Phi_{\a\b\c\d}$ and its
derivatives appears in the right hand sides of \eq{hse1} so the HS
gauge fields become non-zero at second order. Thus, the minimal
bosonic HS gauge theory cannot be consistently truncated to pure
AdS gravity, though it is possible to expand the theory around a
weak gravitational background.

Before proceeding to extract the dynamical field equations to any
order, let us summarize the linearized results \cite{4dv1} (see also \cite
{us3,us4}).
The linearized physical field equations are the
following components of \eq{ee1}, \eq{hse1} and \eq{se1}:

\bea e^{\phantom{(}}_{\a_1\phantom{\bd}\!\!\!}{}^{\bd}\wedge
{\cal R}_{\bd\ad}&=&0 \ , \la{fhse}\w2
e^{\phantom{(}}_{(\a_1\phantom{\bd}\!\!\!}{}^{\bd}\wedge
F^{(1)}_{\a_2\dots\a_{s-1})\bd\ad_1\dots\ad_{s-1}}&=&0\ ,
\la{fee}\w3 (\nabla^2+2)\phi&=&0\ . \la{fse} \eea

The last equation is obtained by using the $m=n=0$ and $m=n=1$
components of \eq{se1}. The remaining components of  \eq{ee1},
\eq{hse1} and \eq{se1} are used in: a) determining the Weyl
tensors $\Phi_{\a(2s)}\equiv \F_{\a_1\dots\a_{2s}}$ ($s=2,4,...$)
in terms of the curvatures; b) solving for the spin connection
$\o_\m^{\a\b}$ in terms of the vierbein $e_\m^{\a\ad}$, and
auxiliary higher spin gauge fields $W_{\m\a(m)\ad(n)}$ for
$|m-n|\ge 2$, $m+n=2s-2$, in terms of the physical fields
$W_{\m\a(s-1)\ad(s-1)}$ ($s=4,6,\dots$); and c) expressing
$\Phi_{\a(m)\ad(n)}$ for $m,n >0$ in terms of derivatives of
$\phi$ and the Weyl tensors. These steps yield

\bea W_{\a\ad,\b_1\dots \b_m\bd_1\dots \bd_n}&=& \ft2{m+1}{\nabla}
W_{\bd_1\bd_2,\a\b_1\dots\b_m \ad\bd_3\dots\bd_n}
+\e_{\ad\bd_1}\ft{2n}{n+1} \Bigg[\ft{n-1}{m+n+2}{\nabla} W_{\bd_2}
{}^{\cd}{}_{,\a\b_1\dots\b_m\cd\bd_3\dots\bd_n} \nn\w2
&&+\ft{n+1}{m+n+2}{\nabla}W_{\phantom{\bd}\!\!\!(\a}{}^{\c}{}_{,\b_1\dots\b_m)
\c\bd_2\dots\bd_n}
-\ft{m}{(m+1)(m+2)}\e_{\phantom{\bd}\!\!\!\a\b_1}{\nabla} W
^{\c\d}{}_{,\c\d\b_2\dots\b_m \bd_2\dots\bd_n}\Bigg] \nn\\ &&+
m\e_{\phantom{\bd}\!\!\!\a\b_1}\x_{\b_2\dots\b_m\ad\bd_1\dots\bd_n}\
,\qquad n>m\geq 0\ ,\quad m+n=6,10,\dots\ , \la{auxsolv1}\w4
\Phi_{\a_1\dots\a_m\ad_1\dots\ad_n}&=& -i
\nabla_{\a_1\ad_1}\Phi_{\a_2\dots\a_m\ad_2\dots\ad_n} \ ,
\la{phisolv1} \eea

where separate symmetrizations $(\a_1...\a_m)$ and
$(\ad_1...\ad_n)$, and similarly $(\b_1...\b_m)$ and
$(\bd_1...\bd_n)$ is understood, the derivatives are Lorentz
covariant and we have used the notation
$W_{\a\ad}=(\s^\m)_{\a\ad}W_\m$ and

\be {\nabla} W_{\a\b,\c_1\dots\c_m\cd_1\dots\cd_n} \equiv
\ft12(\s^{\m\n})_{\a\b} \nabla_\m
W_{\n\c_1\dots\c_m\cd_1\dots\cd_n}\ . \ee

In \eq{auxsolv1} the components
$\x_{\b1\dots\b_{m-1}\bd_1\dots\bd_{n+1}}$ are not
fixed by the generalized torsion constraints. However, from the component form
of the linearized symmetries \eq{lingt},

\be \d
W_{\m\,\a_1\dots\a_m\ad_1\dots\ad_n}=\nabla_\m\e_{\a_1\dots\a_m\ad_1\dots\ad_n}+
m(\s_\m)_{(\a_1}{}^{\bd}\e_{\a_2\dots\a_m)\ad_1\dots\ad_n\bd}+
n(\bar
\s_\m)_{(\ad_1}{}^{\b}\e_{\a_1\dots\a_m\b\ad_2\dots\ad_2)}\ .\ee

it follows that the auxiliary gauge parameter $\e_{\a(m)\ad(n)}$ with
$|m-n|=4,6,...$ acts by shifting the undetermined component
$\x_{\a(m)\ad(n)}$. The undetermined components are
therefore gauge artifacts which can be eliminated by
going to a physical gauge. For example, we can impose

\be
(\s^\m)_{\phantom{\bd}\!\!\!(\a_1}{}^{\bd}W_{\m\,\a_2\dots\a_{m})\ad_1\dots\ad_
{n}\bd}=0\
,\quad m-n=4,6,\dots\ .\la{physgauge}\ee

The first order physical field equations \eq{fhse}--\eq{fse} are
invariant under the remaining generalized Lorentz rotations and
generalized translations with parameters $\e_{\a(s-2)\ad(s)}$ and
$\e_{\a(s-1)\ad(s-1)}$, respectively ($s=2,4,...$). Some of these
gauge symmetries can be used to recover Fronsdal's formulation of
free massless higher spin fields in terms of doubly traceless
symmetric tensors ($s=4,6,...$)

\bea h_{\m_1\dots \m_s}&=&(\s_{\m_2})^{\a_1\ad_1}\cdots
(\s_{\m_2})^{\a_{s-1}\ad_{s-1}}W_{\m_1\a_1\dots\a_{s-1}\ad_1\dots\ad_{s-1}}\
,\\
g^{\m\n}g^{\r\s}h_{\m\n\r\s\l(s-4)}&=&0\ ,\eea

obeying field equations which are invariant under the remaining
translation-like symmetries

\be \d h_{\m_1\dots\m_s}=\nabla_{(\m_1}\e_{\m_2\dots\m_s)}- {\rm
double\ traces}\ .\ee

These symmetries remove the single-traces as well as the time-like
and longitudinal modes, leading to two physical transverse spin
$s$ modes with energy $E_0=s+1$ for $s=4,...$. Together with the
physical scalar with energy $E_0=1$, which is described by
\eq{fse}, and the graviton this results in a spectrum isomorphic
to the symmetric tensor product of two scalar $SO(3,2)$
singletons.

In going beyond first order, each component of the curvature
constraints \eq{cn} and \eq{bn} continues to serve the purpose
it did in the first order analysis. For example, a first order
generalized torsion
constraint, from which an auxiliary gauge field is eliminated in
the leading order, plays the same role in the full theory, though
the elimination procedure must of course be done perturbatively
(see \eq{auxsolv} and \eq{phisolv} below). Similarly, the full
field equations are contained in the same components of \eq{cn}
and \eq {bn} as the first order field equations \eq{fhse} and
\eq{fse}, i.e. they are given by

\bea && e_{\phantom{\bd}\!\!\!\a}{}^{\bd}\wedge {\cal
R}_{\bd\ad}=e_{\phantom{\bd}\!\!\!\a}{}^{\bd}\wedge J_{\bd\ad}\ ,
\la{gfe}\w4 && e_{(\a_1\phantom{\bd}\!\!\!}{}^{\bd}\wedge
F^{(1)}_{\a_2\dots
\a_{s-1})\bd\ad_1\dots\ad_{s-1}}=e_{(\a_1\phantom{\bd}\!\!\!}{}^{\bd}\wedge
J_{\a_2\dots \a_{s-1})\bd\ad_1\dots\ad_{s-1}} \la{hsfe}\w4 &&
(\nabla^2+2)\phi=\nabla^\mu P_\mu -\ft{i}2
(\s^\mu)^{\a\ad}P_{\mu,\a\ad}\ . \la{sfe} \eea
\smallskip

where the two-form $J=J^{(2)}+J^{(3)}+\cdots$ and the one-form
$P=P^{(2)}+P^{(3)}+\cdots$, which contain the interactions, are
given by

\bea
 J &=&-D_E K-(W+K)\star (W+K)-\sum_{n=2}^\infty\sum_{j=0}^{n}
\Bigg( \widehat{A}^{(j)}\star \widehat{A}^{(n-j)}\Bigg)_{Z=0}\ .
\la{jint1}\w3 P&=&\Phi\star \bar \pi(W+K)-(W+K)\star
\Phi+\sum_{n=2}^\infty \sum_{j=1}^{n}
\Bigg(\widehat{\Phi}^{(j)}\star \bar
\pi(\widehat{A}^{(n-j)})-\widehat{A}^{(n-j)}\star
\widehat{\Phi}^{(j)} \Bigg)_{Z=0}\ ,\nn \eea

where $D_E K=dK+E*K+K*E$ and the quantity $K=dx^\m K_\m$ is
defined in \eq{k} and it has the expansion
$K=K^{(2)}+K^{(3)}+\cdots$. The contracted curvatures in the right
hand sides of \eq{gfe} and \eq{hsfe} contain the vierbein
$e_{\mu\, \a\ad}$ and its higher spin generalizations
$W_{\m\,\a_1\dots\a_{s-1}\ad_1\dots\ad_{s-1}}$ and first order
derivatives of the Lorentz connection $\o_{\mu\,\a\b}$ and its
higher spin generalizations
$W_{\mu\,\a_1\dots\a_{s-2}\ad_1\dots\ad_s}$.

The general solutions for the auxiliary fields are modified as

\bea W_{\a\ad,\b_1\dots \b_m\bd_1\dots \bd_n}&=&
\ft2{m+1}\widehat{\nabla} W_{\bd_1\bd_2,\a\b_1\dots\b_m
\ad\bd_3\dots\bd_n} +\e_{\ad\bd_1}\ft{2n}{n+1}
\Bigg[\ft{n-1}{m+n+2}\widehat{\nabla} W_{\bd_2}
{}^{\cd}{}_{,\a\b_1\dots\b_m\cd\bd_3\dots\bd_n} \nn\w2
&&+\ft{n+1}{m+n+2}\widehat{\nabla}W_{\phantom{\bd}\!\!\!(\a}{}^{\c}{}_{,\b_1
\dots\b_m)\c\bd_2
\dots\bd_n}
-\ft{m}{(m+1)(m+2)}\e_{\phantom{\bd}\!\!\!\a\b_1}\widehat{\nabla} W
^{\c\d}{}_{,\c\d\b_2\dots\b_m \bd_2\dots\bd_n}\Bigg]\nn\\ &&+
m\e_{\phantom{\bd}\!\!\!\a\b_1}\x_{\b2\dots\b_m\ad\bd_1\dots\bd_n}\
,\qquad n>m\geq 0\ ,\la{auxsolv}\w4
\Phi_{\a_1\dots\a_m\ad_1\dots\ad_n}&=& -i
\widehat\nabla_{\a_1\ad_1}\Phi_{\a_2\dots\a_m\ad_2\dots\ad_n} \ ,
\la{phisolv} \eea

where the modified covariant derivatives are defined by

\bea && \widehat{\nabla} W_{\a\b,\c_1\dots\c_m\cd_1\dots\cd_n}
=\ft12(\s^{\m\n})_{\a\b} \left(\nabla_\m
W_{\n\c_1\dots\c_m\cd_1\dots\cd_n}-\ft12
J_{\m\n\c_1\dots\c_m\cd_1\dots\cd_n}\right)\ ,\la{nprime}\w3
&&\widehat{\nabla}_{\a_1\ad_1}
\Phi_{\a_2\dots\a_m\ad_2\dots\ad_n} =
\Big(\nabla_{\a_1\ad_1}\Phi_{\a_2\dots\a_m\ad_2\dots\ad_n}
-P_{\a_1\ad_1,\a_2\dots\a_m\ad_2\dots\ad_n}\Big)\ , \la{phisolv2}
\eea

Since $J_{\m\n}$ and $P_\m$ depend on the auxiliary fields, eqs.
\eq{auxsolv} and \eq{phisolv} must be iterated within the
curvature expansion scheme. This leads to explicit expressions of
all auxiliary components of $W_\mu$ and $\Phi$ in terms of the
remaining physical fields \eq{ff} to any desired order. Inserting
these expressions into \eq{gfe}, \eq{hsfe} and \eq{sfe} we find
the physical field equations.

The non-covariant contributions to $J_{\m\n}$ and $P_\m$ from the Lorentz
connection $\o_\m$ must cancel, since the full curvature
constraints as well as the quantities ${\cal R}_{\m\n}$, $F^{(1)}(W)_{\m\n}$
and
$\nabla_\m\F$ are Lorentz covariant. Thus the $\o_\m$ dependence
in $P_\m$ can be dropped immediately, while in $J_{\m\n}$ we first
covariantize $D_E K$ by writing it as

\be D_E
K=iR^{\a\b}\Big(\widehat{A}_\a\star\widehat{A}_\b\Big)_{Z=0} -{\rm
h.c.}+\mbox{non- covariant $\o$ dependence}\ , \ee

where $R_{\a\b}$ is the Riemann curvature two-form. From the
manifest Lorentz invariance it follows that the non-covariant $\o$
dependence in $J$ and $P$ must cancel. Thus:

\bea J&=&-i\left(R^{\a\b}\widehat{A}_\a\star\widehat{A}_\b+
R^{\ad\bd}\widehat{ A}_{\ad}\star\widehat{
A}_{\bd}\right)_{Z=0}-W\star W\nn\w2 &&
-\sum_{n=2}^\infty\sum_{j=0}^{n} \Bigg(
\left(\widehat{e}^{(j)}+\widehat{W}^{(j)}\right)\star
\left(\widehat{e}^{(n-j)}+\widehat{W}^{(n-j)}\right)\Bigg)_{Z=0}\
,\la{jint2}\w3 P&=&\Phi\star \bar\pi(W)-(W)\star \Phi\nn\w2 &&
+\sum_{n=2}^\infty \sum_{j=1}^{n} \Bigg(\widehat{\Phi}^{(j)}\star
\bar
\pi\left(\widehat{e}^{(n-j)}+\widehat{W}^{(n-j)}\right)-\left(\widehat{e}^{(n-
j)} +\widehat{W}^{(n-j)}\right) \star \widehat{\Phi}^{(j)}
\Bigg)_{Z=0}\ ,\la{pint}\eea

where $R_{\ad\bd}=(R_{\a\b})^\dagger$.

It is straightforward to
modify the above formulae such that the gravitational connections
$e$ and $\o$ become linearized around an AdS-like vacuum $\O$
obeying $d\O+\O\star\O=0$. The Lorentz covariant derivative
$\nabla_\m$ then becomes background Lorentz covariant and $e$
splits into AdS background plus a fluctuation part which can be
accounted for by including also spin $s=2$ fluctuations in $W$.

The physical field equations \eq{gfe} and \eq{sfe} are manifestly
invariant under the local Lorentz transformations \eq{dlo} and
\eq{dlo2} and \eq{dlo3}. They are also invariant under
diffeomorphisms and HS gauge symmetries. In general, a set of
integrable constraints

\be R^I\equiv d C^I+F^I(C^J)=0\ ,\quad {\partial F^I\over \partial
C^J}\wedge F^J=0\ ,\ee

where $C^I$ are $n_I$-forms, have the gauge symmetry

\be \d C^I=\left\{\ba{ll}d\e^I+ {\partial F^I\over \partial
C^J}\wedge \e^J\ ,& n_I\geq 1\w3 {\partial F^I\over \partial
C^J}\wedge \e^J\ ,& n_I=0\ea\right.\ee

Diffeomorphisms, $\d W^I=\{d,i_\x\} W^I$ where $\x=dx^\m \x_\m$,
are incorporated into the gauge group as field dependent
transformations with gauge parameter $\e^I=i_\x W^I$. To apply
these formulae to the full constraints \eq{cn} and \eq{bn} we set
$C^I=(e,\o,W;\F)$. We then find that the gauge transformations
associated with $e$ and $W$ are

\bea \d (e+\o+W)&=&\nabla \e +
[e+W,\e]_{\star}+\sum_{n=2}^\infty\sum_{j=0}^{n} \Bigg(\left[
\widehat{e}^{(j)}+\widehat{W}^{(j)},
\widehat{\e}^{(n-j)}\right]_{\star}\Bigg)_{Z=0}\ ,\la{fullgt1}\w3 \d
\F &=&\Phi\star \bar\pi(\e)-\e\star \Phi+\sum_{n=2}^\infty
\sum_{j=1}^{n} \Bigg(\widehat{\Phi}^{(j)}\star
\bar\pi\left(\widehat{\e}^{(n-j)}-\widehat{\e}^{(n- j)}\star
\widehat{\Phi}^{(j)}\right) \Bigg)_{Z=0}\ ,\la{fullgt2}\eea

where $\e$ is either an ordinary translation,
$\e_{\a\ad}P^{\a\ad}$ or a HS gauge parameter, $\e_{\a(m)\ad(n)}$,
$m+n=6,10,...$. On general grounds, the above transformations must
agree with the form of the residual $Z$-space transformations
discussed at the end of Section 2. As in the first order, the
auxiliary gauge symmetries with parameters $\e_{\a(m)\ad(n)}$,
$|m-n|=4,6,...$ act by curvature corrected shifts of the
$\x$-components in \eq{auxsolv}, which can be used to impose a
physical gauge, such as \eq{physgauge}. In general, there may of
course be other physical gauge choices which lead to further
simplification. The remaining gauge symmetries, i.e. the local
translations and Lorentz transformations and their higher spin
generalizations with $|m-n|=0,2$, respectively, act as symmetries
of the dynamical field equations in \eq{gfe}, \eq{hsfe} and
\eq{sfe}.


\section{Closer Look at the Quadratic Terms in The Field Equations}


In this section, we collect the results of the previous section
for the important special case of the quadratic terms in the field
equations. The field equations up to quadratic order in weak field
approximation are given by

\bea (\nabla^2+2)\,\phi &=& \left( \nabla^\mu P^{(2)}_\mu -\ft{i}2
(\s^\mu)^{\a\ad} {\partial\over \partial y^\a}
{\partial\over\partial {\yb}^{\ad}}\, P^{(2)}_\mu\right)_{Y=0}\ ,
\la{sfe3}\w4 \left(\s^{\m\n\r}\right)_\a^{\bd}\,{\cal
R}_{\n\r\,\bd\cd}&=& \left(\s^{\m\n\r}\right)_\a^{\bd}\, \left(
{\partial \over\partial {\yb}^{\bd}} {\partial \over \partial
{\yb}^{\cd}}J^{(2)}_{\n\r}\right)_{Y=0}\ , \la{efe3}\w4
\left(\s^{\m\n\r}\right)_{(\a_1}^{\bd}\,
F^{(1)}_{\n\r\,\a_2...\a_{s-1})\bd \cd \ad_2 ...\ad_{s-1}}&=&
\left(\s^{\m\n\r}\right)_{(\a_1}^{\bd}\,\left( {\partial \over
\partial y^{\a_2}}\cdots {\partial\over \partial y^{\a_{s-1})}}
{\partial\over
\partial {\yb}^{\bd}}\cdots {\partial\over \partial {\yb}^{\ad_{s-1}
\phantom{\bd}\!\!\!}}\, J^{(2)}_{\n\r}\right)_{Y=0} \la{hsfe3}
\eea

where  ${\cal R}_{\m\n\ad\bd} \equiv  F_{\m\n\ad\bd}$ is the
(self-dual part of) the $AdS_4$ valued Riemann curvature, while
the curvature associated with spin $s=4,6,8,..$ fields is defined
as

\bea F^{(1)}_{\n\r\,\a_2...\a_{s-1}\bd \cd \ad_2 ...\ad_{s-1}} &=&
2\nabla_{[\n }W_{\r ]\a_2...\a_{s-1}\bd\cd\ad_2...\ad_{s-1}}
\nn\w2 &&-(s-2)(\s_{\n\r}\s_\m)_{(\a_2}{}^\d\,
W_{\m\,\a_3...\a_{s-1})\d\bd\cd{\dot\d}\ad_2...\ad_{s-1}} \nn\w2
&& -s (\s_\m\s_{\n\r})_{(\bd}{}^\c\,
W_{\m\,\c\a_2\a_3...\a_{s-1}\cd\ad_2...\ad_{s-1})}\ . \la{f1} \eea

The covariant derivatives in \eq{sfe3} and \eq{f1} are with
respect to the Lorentz connection $\o$. Further definitions are

\bea P_\m^{(2)} &=& \Phi\star\bar{\pi}(W_\m)- W_\m\star \Phi
\nn\w2 && +\Bigg(\Phi\star\bar{\pi}({\widehat e}_\m{}^{(1)})
-\widehat{e}_\m^{(1)}\star \Phi+\widehat{\Phi}^{(2)}\star
\bar{\pi}(e_\m)-e_\m\star\widehat{\Phi}^{(2)}\Bigg)_{Z=0}
\la{p2}\w4 J_{\m\n}^{(2)}&=& -\Bigg[
\Bigg(\widehat{e}_\m^{(1)}\star \widehat{e}_\n^{(1)}+
\{e_\m,\widehat{e}_\n^{(2)}\}_{\star} +\{e_\m,
\widehat{W}_\n^{(1)}\}_{\star}+ \{W_\m,
\widehat{e}_\n^{(1)}\}_{\star} \Bigg)_{Z=0} \nn\w2 && +\Bigg(i
R_{\m\n}{}^{\a\b}\widehat{A}^{(1)}_\a\star\widehat{A}^{(1)}_\b
+\mbox{h.c.} \Bigg)_ {Z=0} +W_\m\star W_\n \Bigg] + \Bigg[\m
\leftrightarrow \n\Bigg]\ , \la{j2} \eea

where the hatted quantities are defined as

\bea
\hA^{(1)}_\a  &=&   -{i \over 2} z_\a \int_0^1 tdt~ b_1
\F(-tz,\yb)\kappa(tz,y)\ , \la{a1}\w4
\widehat{A}^{(2)}_\a &=& z_\a \int_0^1 tdt\left(
\widehat{A}^{(1)\b}\star \widehat{A}^{(1)}_\b\right)_{z\ra
tz,\zb\ra t\zb} \la{a2}\w2 &&+\zb^{\bd}\int_0^1 tdt \left[
\widehat{A}^{(1)}_{\a},\widehat{A}^{(1)}_{\bd}\right]_{z\ra
tz,\zb\ra t\zb} \nn\w4
\widehat{W}^{(1)}_\mu &=&\widehat{L}^{(1)}(W_\mu) \la{w1}\w2
&=& -i\int_0^1{dt\over t} \Bigg(\left[{\partial W_\m\over \partial
y^\a},\hA^{\a(1)}\right]_* +\left[\hA^{\ad (1)} ,{\partial
W_\m\over \partial \yb^{\ad}}\right]_* \Bigg)_{z\ra tz,\zb\ra
t\zb} \nn\w4
{\widehat\Phi}^{(2)}&=& z^\a\int_0^1dt\left[\Phi\star
{\bar\pi}(\hA_\a^{(1)}) - \hA_\a^{(1)}\star \Phi
\right]_{t\rightarrow tz, \zb \rightarrow t\zb} \la{phi2}\w3
&&+{\zb}^{\ad}\int_0^1 dt \left[\Phi\star {\pi}({\widehat
A}_{\ad}^{(1)}) - \hA_{\ad}^{(1)}\star \Phi \right]_{t\rightarrow
tz, \zb \rightarrow t\zb} \nn\w4
\widehat{e}^{(1)}_\mu &=&\widehat{L}^{(1)}(e_\mu) \la{e1}\w2
&=& -i e_\m^{\a\ad} \int_0^1 {dt\over t} \Bigg(
\left[{\yb}_{\ad},\hA_\a^{(1)}\right]_*
+\left[\hA_{\ad}^{(1)},y_\a\right]_* \Bigg)_{z\ra tz,\zb \ra
t\zb}\ , \nn\w4
\widehat{e}^{(2)}_\mu&=&(\widehat{L}^{(2)}
+\widehat{L}^{(1)}\widehat{L}^{(1)})(e_\mu) \la{e2}\w3 &=&-i
e_\m^{\a\ad} \int_0^1 {dt\over t} \Bigg(
\left[{\yb}_{\ad},\hA_\a^{(2)}\right]_* +\left[{\widehat
A}_{\ad}^{(2)},y_\a\right]_* \Bigg)_{z\ra tz,\zb \ra t\zb}
\nn\w3
&& -e_\m^{\a\ad} \int_0^1 {dt\over t} \int_0^1 {dt'\over t'}
\Bigg[ \hA^{\b(1)}\star \left({\partial\over\partial
z^\b}-{\partial\over\partial y^\b}\right) \Big(
\left[{\yb}^{\phantom{()}}_{\ad},\hA_\a^{(1)}\right]_*
                     +\left[\hA_{\ad}^{(1)},y^{\phantom{()}}_\a\right]_*
                     \Big)_{z\ra t'z,\zb \ra t'\zb}
\nn\w3
&& \qquad\qquad\qquad\qquad + \hA^{\bd(1)}\star
\left({\partial\over\partial \zb^{\bd}}+{\partial\over\partial
\yb^{\bd}}\right) \left(
\left[{\yb}^{\phantom{()}}_{\ad},\hA_\a^{(1)}\right]_*
+\left[\hA_{\ad}^{(1)},y^{\phantom{()}}_\a\right]_*
                     \right)_{z\ra t'z,\zb \ra t'\zb}
\nn\w3
&& \qquad\qquad\quad + \left({\partial\over\partial
z^\b}+{\partial\over\partial y^\b}\right)\left(
\left[{\yb}^{\phantom{()}}_{\ad},\hA_\a^{(1)}\right]_*
                     +\left[\hA_{\ad}^{(1)},y^{\phantom{()}}_\a\right]_*
                     \right)_{z\ra t'z,\zb \ra t'\zb} \star \hA^{\b (1)}
\nn\w3
&& \qquad\qquad\quad + \left({\partial\over\partial
\zb^{\bd}}-{\partial\over\partial \yb^{\bd}}\right)\left(
\left[{\yb}^{\phantom{()}}_{\ad},\hA_\a^{(1)}\right]_*
                     +\left[\hA_{\ad}^{(1)},y^{\phantom{()}}_\a\right]_*
                     \right)_{z\ra t'z,\zb \ra t'\zb} \star \hA^{\bd(1)}
                     \Bigg]_{z\ra tz,\zb \ra t\zb}\ . \nn
\eea

where the constant $b_1$ in \eq{a1} is defined by \eq{odd} and
\eq{be1}. In the above formulae, the replacement
$(z,\zb)\ra(tz,t\zb)$ in a quantity must be performed {\it after}
the quantity has been written in Weyl ordered form i.e. after the
$\star$ products defining the quantity has been performed.

A closer look at the field equations presented above shows that:

\begin{itemize}

\item{} The $z$-dependence of all the fields involved is either a
simple dependence which is explicitly shown or it arises through
dressing via the action of the ${\widehat L}$ operator. All in
all, the above results are explicit and the remaining task is the
evaluation of certain star products (involving finite order
differentiations in $y$ and $z$) and some elementary parameter
integrals. The results of these computations will be given
elsewhere \cite{wip}.

\item{} The field equation \eq{efe3} for the graviton can be rewritten
as
\footnote{We have set the AdS radius $R=1$ but it is
straightforward to re-introduce $R$ by dimensional analysis in
which the master 0-form and the master 1-form fields are
dimensionless.}

\be R_{\m\n}(\o)-g_{\m\n} =  \left[
(\s_\m{}^\l)^{\a\b}\,\left({\partial \over \partial
y^\a}{\partial\over \partial y^\b}\,J^{(2)}_{\l\n} \right)_{Y=0} +
( \m \leftrightarrow \n ) + \mbox{h.c.}\right] \ ,\la{stress}\ee

where $R_{\m\n}(\o)$ is the Ricci tensor obtained from the Riemann
tensor associated with the Lorentz connection $\o_\m$. It is
important to note that this connection contains torsion as can be
seen from \eq{auxsolv} and \eq{nprime} which for $m=0, n=2$ give

\be \o_\m{}^{ab} = \o_\m{}^{ab}(e)+\k_\m{}^{ab}\ , \ee

where $\k_\m{}^{ab}$ is the con-torsion tensor related to the
torsion tensor $T_{\m\n}{}^a$ as

\be \k_\m{}^{ab}=T_\m{}^{ab}-T_\m{}^{ba}+T^{ab}{}_\m\ , \ee

where

\be T_{\m\n}{}^a= \left(\s^a\right)_{\a\bd}\,\left( {\partial
\over\partial y_\a} {\partial \over \partial
{\yb}_{\bd}}J^{(2)}_{\m\n}\right)_{Y=0}\ . \ee

\item{} In the final form of the field equations, the auxiliary
fields arise both in the master scalar field $\Phi$ as well as the
master gauge field $A_\m$. Their solution given in \eq{phisolv}
and \eq{auxsolv} in terms of the physical fields are to be used in
these equations. It is through the elimination of these auxiliary
fields that higher derivative interactions arise. In particular,
in a given spin sector, the auxiliary fields are $W_{\m
\a_1...\a_k\ad_{k+1}...\ad_{2s-2}}$ with $k=0,1,...,s/2$ and they
are related to the physical fields $W_{\m\a(s-1)\ad(s-1)}$
schematically as

\be W_{\m \a(m)\ad(n)}\ \sim \
\partial^{|m-n|/2}\, W_{\m\a(s-1)\ad(s-1)}\ ,\qquad m+n=2s-2\ .\ee

Similarly, the components $\Phi_{\a(m)\ad(n)}$of the master scalar
field are related to the Weyl tensors which are purely chiral,
their derivatives as well as the derivatives of the scalar as
(taking $m>n$ without loss of generality)

\bea \Phi_{\a(m)\ad(m)} \ &\sim&\ \partial^m\,\phi\ ,\nn\w2
\Phi_{\a(m)\ad(n)}\ &\sim& \ \partial^{(m-n)/2}\,\Phi_{\a(m-n)}\
,\qquad m-n=0\, {\rm mod} \,4\ .\eea

The preliminary results of \cite{wip} indicate that the
contribution from the physical scalar field to the stress energy
tensor in \eq{stress} involves derivatives of the scalar to
arbitrary order period. Whether the higher orders can be absorbed
into a field redefinition remains to be seen.

\item{} The field equations \eq{sfe} and \eq{hsfe3} are invariant
under generalized Lorentz rotations and translations. Their
action, including the compensating gauge transformations due to
\eq{physgauge}, is given by \eq{fullgt1} and \eq{fullgt2}, where
one of course retains only terms up to quadratic in order in $\e$,
$W$ and $\Phi$.

\end{itemize}


\section{Discussion}


In this paper we have studied the structure of a minimal bosonic
HS gauge theory in four dimensions. We have presented in
considerable detail an iterative procedure for determining the
field equations to any desired order in weak fields, which are the
physical scalar and the higher spin gauge fields as well as
arbitrary derivatives of the physical fields, including the spin
$2$ Weyl tensor and its derivatives. We note that the vierbein is
not expanded around any fixed background and the field equations
are manifestly reparametrization and local Lorentz invariant.

We emphasize that the building blocks of the theory are the master
0-form ${\hat\Phi}$ containing the physical scalar, the Weyl
tensors and derivatives, and a master 1-form ${\hat A}$ containing
the physical spin $s=2,4,...,\infty$ fields and auxiliary fields
which can be solved for in terms of the physical fields and their
derivatives.  The central equation that describes the theory is
(see Section 2)

\be d\hF+\hA\star \hF-\hF\star \bar{\pi}(\hA)=0 \ .\la{1}\ee

The integrability of this equation naturally leads to the second
master equation (modulo the generalization involving $\cV(X)$
discussed in Section 2)

\be  d\hA+\hA\star\hA= \ft{i}4 dz^\a\wedge dz_\a\, \cV(\hF\star
\k) -\mbox{h.c.}\ ,\la{2} \ee

which together with \eq{1} form a complete set of constraints
defining the theory. For weak $\hF$, eqs. \eq{1} and \eq{2} are
equivalent to the constraints \eq{cn} and \eq{bn} in ordinary
spacetime, which have been the main focus of the discussion in
this paper. It is worth pointing out, however, that \eq{1} and
\eq{2} imply that $\hA_\m$ is locally a pure gauge field. Thus, in
a neighborhood of a spacetime point, $p$ say, \eq{1} and \eq{2},
and therefore \eq{cn} and \eq{bn}, are equivalent to $\widehat
F_{\a\ad}|_p=0$, $(\widehat
F_{\a\b}+\ft{i}2\e_{\a\b}\cV(\hF\star\k))|_p=0$ and $\hF |_p=0$.
The use of this in the context of the AdS/HS gauge theory
correspondence remains to be examined.

The master scalar field is essential for the consistency of the
theory in that it contains the physical scalar field which is
required for a physical spectrum that fits into an UIR of the HS
symmetry group that underlies the theory. At the same time, the
presence of the master scalar field complicates the problem of
finding an action. Indeed, in the absence of the scalar, one can
write down an action of the type $S=\int d^4x {\rm Tr}
F\star\wedge F$, where Tr is the trace over the HS algebra
\cite{vr2}. However, since this action does not contain the
physical scalar field it does not describe a consistent HS gauge
theory. This problem does not arise in 3D, where actions for
higher spin gauge theories including their matter couplings have
been constructed; see, for example, \cite{blencowe,vr2}.

The master field equations \eq{1} and \eq{2} are deceptively
simple looking. They embody equations that contain curvatures of
the physical fields and their derivatives to arbitrarily high
order, which can be obtained in a weak field expansion by the
iterative procedure explained in the paper. A glance at the
resulting equations up to second order in weak fields given in
eqs. \eq{sfe3} and \eq{hsfe3} reveal that the detailed form of the
equations is very complicated. Considerable may be achieved by
working in a first order formalism where the auxiliary fields are
treated as independent fields. The fact that there exist composite
operators on the boundary corresponding to the auxiliary fields
\cite{vcurrents} is encouraging in this respect.

The results of this paper can easily be carried over to the HS
gauge theory extension of the $D=4, {\cN}=8$ supergravity. This
theory admits the minimal bosonic theory as a consistent
truncation. A search for a holographic description of the $\cN=8$
theory is the most interesting one as it is naturally related to
$M$ theory on $AdS_4\times S^7$. Nonetheless, some basic aspects
of holography in this context can already be glimpsed from a study
of the minimal bosonic model. In that case, a relevant boundary
field theory to consider is scalar singleton $W$ in the adjoint
representation of $SU(N)$ for large $N$. Bilinear operators
corresponding to the physical spectrum of the bulk HS gauge theory
can be built out of two such singletons and their derivatives. The
operators corresponding to spin $s=2,4,...$ fields are conserved
tensors, while the physical scalar couples to $:{\rm tr}W^2:$.

The free singleton theory also contains composite operators which
are cubic or higher order in the singleton field $W$ and that
couple to massive bulk fields. It has been proposed in \cite{holo}
that in the $\cN=8$ theory the massless bulk fields decouple
consistently from the massive bulk fields in the leading order in
the $1/N$ expansion. This is crucial for the massless HS theory to
make sense as a limit of M theory since there is no mass-gap
between the massless and the massive fields. The truncation to the
massless sector can be tested since it implies that there exists a
choice of interaction $\cV(X)$ given in \eq{2} such that bulk
theory reproduces the correlators of the bilinear operators on the
boundary. Moreover, this test can be performed at the level of the
massless minimal bosonic theory \cite{wip}, which is a consistent
truncation of the massless $\cN=8$ theory.

\begin{center} {\bf \Large Acknowledgements} \end{center}

P.S. is thankful to U. Danielsson, J. Engquist, F. Kristiansson
and P. Rajan for discussions. We also thank E. Witten for
stimulating  correspondence.

\pagebreak


\end{document}